\documentclass[twocolappendix,iop,apj]{emulateapj}
\usepackage{amsfonts,amsmath,graphicx,natbib,url,hyperref,nicefrac}
\hypersetup{colorlinks=true, urlcolor=blue, citecolor=blue}

\usepackage[usenames,dvipsnames]{color}
\usepackage{comment}

\newcommand{\jp}{\ensuremath{{P}_k}}
\newcommand{\jq}{\ensuremath{{Q}_k}}

\shorttitle{Jet Model I} 
\shortauthors{Duffell \& Laskar}

\begin{document}

\title{On the Deceleration and Spreading of Relativistic Jets I:  Jet Dynamics}

\def\ucb{1}
\def\nrao{2}

\author{Paul C. Duffell\altaffilmark{\ucb} and Tanmoy Laskar\altaffilmark{\ucb,\nrao}}
\altaffiltext{\ucb}{Department of Astronomy and Theoretical Astrophysics Center, University of California, Berkeley, CA 94720}
\altaffiltext{\nrao}{Jansky Fellow, National Radio Astronomy Observatory, 520 Edgemont Road, Charlottesville, VA 22903, USA}
\email{duffell@berkeley.edu}

\begin{abstract}

Jet breaks in gamma ray burst (GRB) afterglows provide a direct probe of their collimation angle.  Modeling a jet break requires an understanding of the ``jet spreading" process, whereby the jet transitions from a collimated outflow into the spherical Sedov-Taylor solution at late times.  Currently, direct numerical calculations are the most accurate way to capture the deceleration and spreading process, as analytical models have previously given inaccurate descriptions of the dynamics.  Here (in paper I) we present a new, semi-analytical model built empirically by performing relativistic numerical jet calculations and measuring the relationship between Lorentz factor and opening angle.  We then calculate the Lorentz factor and jet opening angle as a function of shock radius and compare to the numerical solutions.  Our analytic model provides an efficient means of computing synthetic GRB afterglow light curves and spectra, which is the focus of paper II.

\end{abstract}


\section{Introduction} \label{sec:intro}

One of the most valuable tools in measuring GRB jet parameters has been careful analysis of the afterglow emission, which is observed immediately after the prompt emission (seconds), and is detectable for long durations in time (months), at wavelengths across the electromagnetic spectrum.  X-Ray observations from Swift and Chandra have enabled direct measurements of jet properties \citep{1999ApJ...519L..17S, 2006ApJ...642..389N,2009ApJ...698...43R, cfh+10,cfh+11,2015ApJ...814....1L}.

One important feature of the afterglow light curve is the jet break.  Roughly speaking, this occurs around the time the opening angle of the jet is of order the inverse of the jet Lorentz factor ($\Gamma \theta_j \sim 1$).  At this time, the light curve steepens, for two reasons.  First, due to relativistic beaming, at this time it becomes possible for the observer to see the edge of the jet, so the decay of the light curve becomes faster than it would if the GRB were a spherical outflow.  Secondly, around this time the center of the jet comes into causal contact with the edge, and the dynamics no longer proceed as if the jet were spherically symmetric.  The jet begins to spread, causing it to sweep up more mass and decelerate faster, until the flow becomes spherical.

A measurement of the jet break time makes it possible to constrain the initial opening angle ($\theta_0$) of the jet, allowing for a determination of the true energy scale. 
Additionally, inferred rates depend on the fraction of jets which are pointed at us, which is determined by the jet opening angle.  Finally, knowing how collimated the jet is may help to constrain properties of the progenitor.

As mentioned above, the jet break is due to two different effects, one observational and one dynamical.  The observational effect is straightforward to model, but the dynamical effect is more complicated, as it involves knowing how relativistic jets spread in time, which is a nontrivial question.  Intermediate asymptotic solutions have been found for this spreading phase \citep{2007arXiv0704.3081G, 2015ApJ...815..100K}, but it is still not clear how relevant these solutions are for realistic jets.

Currently, the best way to model the jet spreading phase is through direct hydrodynamical calculations \citep{2009ApJ...698.1261Z, 2010ApJ...722..235V, 2010MNRAS.403..300V, 2011ApJ...733L..37V, 2013ApJ...767..141V, 2015ApJ...806..205D}.  By numerically integrating the relativistic fluid equations for a jet with a given initial opening angle, and post-processing this solution with a sufficiently detailed radiative transfer calculation, an accurate GRB afterglow light curve can be obtained.  Additionally, this process can be made more efficient by exploiting the inherent scale invariance of both the hydrodynamical equations and the synchrotron emission spectrum \citep{2012ApJ...747L..30V}.  By performing only a handful of numerical calculations for different initial jet opening angles, one spans the entire parameter space of solutions \citep{2012ApJ...749...44V, 2012ApJ...751...57D}, making it possible to invert the problem and find best-fit parameters for a given GRB afterglow \citep{2015ApJ...799....3R, 2015ApJ...806...15Z}.

However, reliance on numerical models, especially for model-fitting, can be computationally intensive, and force users to go along with the assumptions used in post-processing the simulations to derive light curves. Analytic models to describe the jet evolution can be more versatile, and if easily implemented, they are extremely useful.
\cite{1999ApJ...525..737R} modeled the jet evolution as a coupled system of ordinary differential equations (ODEs) for the jet Lorentz factor and opening angle.  Unfortunately, this model has been shown to be inaccurate.  One of the signatures of this model is an exponential spreading phase, where the shock spreads rapidly sideways almost immediately after the entire jet is in causal contact.  Such a rapid spreading phase has not been seen in numerical studies \citep{2009ApJ...698.1261Z, 2010ApJ...722..235V, 2012ApJ...751..155V}.
\cite{2012MNRAS.421..522L} suggested that this discrepancy was due to an inaccurate criterion for when the jet begins to spread,
and proposed the new criterion, $\Gamma \sim 1/\theta_j^{1/2}$.
\cite{2012MNRAS.421..570G} found that the discrepancy between analytical and numerical results could be resolved without such drastic changes to the model, and proposed several model improvements.  Various new models were compared with a numerical calculation by \cite{2012ApJ...751...57D}, showing that all of these models were reasonably consistent with this numerical result.  However, no clear best-fit model was determined in that study, and the results were only compared with a single numerical calculation, with a fairly wide initial opening angle ($\theta_0 = 0.2$).

In the present study, we build a new semi-analytical model, 
building upon 
\citet[][hereafter R99]{1999ApJ...525..737R} and \citet[][hereafter GP12]{2012MNRAS.421..570G}.  However, rather than proposing several models and comparing them with a numerical calculation, this study proceeds in the opposite direction.  
We perform several numerical calculations of jet spreading using a range of jet properties and circumburst density profiles. We then provide a unique analytic model, with dimensionless constants calibrated to match the numerical calculations. The underlying physics of our model is consistent with most of the assumptions of R99 and GP12.

We begin by describing the models of R99 and GP12 in more detail, and incorporating our changes to the equations (Section \ref{sec:anly}).  Next, the numerical method and initial conditions are specified (Section \ref{sec:numerics}), after which best-fit parameters are found by matching our models to full relativistic hydrodynamics calculations (Section \ref{sec:dyn}).  Results are summarized in Section \ref{sec:disc}.

\section{Analytical Jet Model} \label{sec:anly}

\subsection{ Four-Velocity }

The most important improvement of GP12 was to extend the model of R99 to nonrelativistic velocities.  To do so, the evolution of the jet is expressed in terms of its four-velocity, $u = \Gamma \beta$, as opposed to its Lorentz factor, $\Gamma$.

R99 described jet evolution based on energy and momentum conservation arguments of \cite{1993ApJ...418L...5P}.  According to these arguments, the jet four-velocity is given by

\begin{equation}
u = { \Gamma_0 \over \sqrt{ 1 + 2 \Gamma_0 f + f^2 } },
\label{eqn:vel1}
\end{equation}
where $\Gamma_0$ is the initial Lorentz factor and the quantity $f$ is given by the ratio of swept-up mass, $m$, to the ejecta mass, $M_0$:
\begin{equation}
f = m/M_0.
\end{equation}

However, GP12 noted that momentum conservation cannot be used to determine the jet velocity when the flow becomes spherical, as it is based on conservation of linear momentum, which is zero for a spherical outflow.  At late times, the shock velocity can be estimated using conservation of energy, assuming the total energy is given by its initial value $E = \Gamma_0 M_0$:

\begin{equation}
u \sim \sqrt{E/m} \sim \sqrt{ \Gamma_0 M_0 / m } \sim \sqrt{ \Gamma_0 / f }.
\end{equation}

This asymptotic form suggests that (\ref{eqn:vel1}) should be modified.  This can be done by adding another factor to compensate in this asymptotic limit.  Four-velocity in our model is given by

\begin{equation}
u = { \Gamma_0 \over \sqrt{ 1 + 2 \Gamma_0 f + f^2 } }\sqrt{ 1 + f/\Gamma_0 },
\label{eqn:vel2}
\end{equation}
which now agrees with the Sedov scaling in the spherical, nonrelativistic regime.

\subsection{ Entrained Mass }

A formula must be specified for the mass swept up by the jet, to determine the value of $f$.  R99 suggested


\begin{equation}
f = {1 \over M_0} \int_0^r \rho(r') \Omega(r') r'^2 dr',
\end{equation}
where $\Omega(r)$ is the solid angle subtended by the jet when it has reached radius $r$.  GP12 called this the ``trumpet model", as it suggests that as the jet spreads, it does not entrain any of the mass that it overtakes by spreading, and a diagram for what mass is swept up by the jet takes on a flared, trumpet-like shape.

GP12 proposed an alternate ``conical model", which is more consistent with what is seen in numerical studies.  In this case, the mass swept up by the jet consists of all mass between the shock front and $r = 0$ within the cone subtended by the opening angle $\theta_j$.  In this scenario, $\Omega(r)$ can be pulled out of the integral, which no longer depends on the history of the jet spreading process:

\begin{equation}
f = {\Omega \over M_0} \int_0^r \rho(r') r'^2 dr'.
\end{equation}

If $\rho = A r^{-k}$, then

\begin{equation}
f = {\Omega A r^{3-k} \over (3-k) M_0}.
\label{eqn:f2}
\end{equation}

In practice, we have found the ``conical model" to be more accurate than the ``trumpet model".  This is consistent with what is seen in numerical studies; while the jet energy is confined within the opening angle $\theta_j$, there is also a bow shock at larger angles carrying an insignificant fraction of the jet energy, yet still entraining mass from the surrounding medium.  The spreading of the jet is a process whereby the energy is redistributed into this bow shock, which has already entrained all the mass between the shock front and the origin at $r=0$.  Thus, we can make the assumption that the amount of mass entrained in the jet does not depend on the history of $\theta_j$ with time, only on its value at the given time.

\subsection{Dynamics of Spreading}

The equations (\ref{eqn:vel2}) and (\ref{eqn:f2}) can be closed by giving $\theta_j$ as a function of the Lorentz factor.  R99 assumed that the jet expanded as a sound wave, starting at $\theta = \theta_0$ and moving outward at the sound speed in the observer frame:

\begin{equation}
\theta_j = \theta_0 + {t_{\rm co} \over \sqrt{3} r},
\end{equation}
where
\begin{equation}
t_{\rm co} = \int dt / \Gamma(t).
\end{equation}

GP12 extended this model by accounting for the curvature of the shock front:

\begin{equation}
{d\theta \over d {\rm ln} r} \sim {1 \over \Gamma^{1+a} \theta^{a}},
\end{equation}
where the parameter $a$ is either $0$ or $1$, depending how the shape of the shock front is modeled.  It was found in that study that both choices were reasonably consistent with the numerical result with which they were compared.


To determine how fast a sound wave can propagate along an expanding surface, one can treat it much like a cosmology problem.  The Minkowski metric is written down in 2+1 dimensions (polar coordinates), and restricted to a 1+1 dimensional surface which is an expanding circle with velocity $dr/dt = v$ (and Lorentz factor $\Gamma$):

\begin{equation}
ds^2 = -dt^2 + dr^2 + r^2 d\theta^2
\end{equation}
\begin{equation}
 = -{1 \over \Gamma^2} dt^2 + r^2(t) d\theta^2
\end{equation}
\begin{equation}
 = -{1 \over u^2} dr^2 + r^2 d\theta^2
\end{equation}
where $u = \Gamma v$ is the four velocity.  Assuming that light rays propagate along null geodesics with $ds^2 = 0$, this would imply the following relationship, which is consistent with GP12:

\begin{equation}
{d \theta \over d {\rm ln} r} = {1 \over u}.~~~~\text{(GP12, $a=0$)}
\end{equation}

This equation represents the fastest possible lateral motion allowed by causality.  In reality, it should be modified because the jet does not spread at the speed of light.  R99 assumed that the jet spreads at the speed of sound, but it may take several sound-crossing times for the jet energy to redistribute itself, so the spreading velocity is left as a free parameter, $c_j$:

\begin{equation}
{d \theta_j \over d {\rm ln} r} = {c_j \over u}.
\label{eqn:theta2}
\end{equation}

It is possible to derive a closed-form relationship between $u$ and $\theta$ during the early spreading phase, using (\ref{eqn:vel2}), (\ref{eqn:f2}) and (\ref{eqn:theta2}).  During the spreading phase, $2 \Gamma_0 f \gg 1$, so that (\ref{eqn:vel2}) can be approximated as
\begin{equation}
u = \sqrt{ \Gamma_0 \over 2 f }.
\end{equation}
Taking the exterior derivative of both sides, after algebra one obtains,
\begin{equation}
d {\rm ln} u = - \frac12 d {\rm ln} f.
\end{equation}
Using (\ref{eqn:f2}) for $f$,
\begin{equation}
d {\rm ln} u = - \frac12 ( d {\rm ln} \Omega + (3-k) d {\rm ln} r ).
\end{equation}
Assuming the jet is still narrow, and using the small angle approximation $\Omega = \pi \theta_j^2$, one obtains $d {\rm ln} \Omega = 2 d {\rm ln} \theta_j$, and
\begin{equation}
d {\rm ln} u = - d {\rm ln} \theta_j - \frac12 (3-k) d {\rm ln} r.
\end{equation}
Combining with (\ref{eqn:theta2}) to eliminate $r$:
\begin{equation}
d {\rm ln} u = - d {\rm ln} \theta_j - {(3-k) u \over 2 c_j} d \theta_j.
\end{equation}
For compactness, we define the dimensionless constant $\jp \equiv (3-k)/(2 c_j)$.  The relationship between four-velocity and opening angle is therefore 

\begin{equation}
{d {\rm ln} u \over d {\rm ln} \theta_j} = - 1 - \jp \theta_j u.
\label{eqn:utheta}
\end{equation}

If the spreading is very fast, so that $\jp$ is very small, and the last term is neglected, one is left with $u \theta_j = $ constant, reminiscent of the R99 exponential spreading phase in which $\Gamma \sim 1/\theta_j$.  Such a phase can occur, if the product $\jp \theta_j u$ becomes small enough.  In practice, for typical initial opening angles such a phase does not occur, as was noted by GP12 (consistent with what has been seen in numerical studies).

Equation (\ref{eqn:utheta}) can be solved by re-writing it as an equation for $u \theta_j$:

\begin{equation}
{d {\rm ln} (u \theta_j) \over d \rm ln \theta_j} = - \jp \theta_j u,
\label{eqn:ut1}
\end{equation}
with the solution:

\begin{equation}
u = {1 / \theta_j \over \jq + \jp \rm ln (\theta_j/\theta_0) },
\label{eqn:ut2}
\end{equation}
where $\theta_0$ is the initial opening angle of the jet, and $\jq$ is an arbitrary integration constant, which will be fitted to numerical results.  $\jq$ determines the time when the jet begins to spread; this occurs when $u = 1/(\jq \theta_0)$.  Before this time, the opening angle of the jet remains at its initial value, $\theta_j = \theta_0$.

For larger opening angles, using $\Omega = 4\pi {\rm sin}^2 (\theta_j/2)$ instead of $\Omega = \pi \theta_j^2$, the extension of (\ref{eqn:ut1}) is

\begin{equation}
{d {\rm ln} (u {\rm sin}(\theta_j/2) ) \over d \rm ln~sin (\theta_j/2)} = - \jp u (2 {\rm tan} (\theta_j/2) ),
\end{equation}
resulting in the solution

\begin{equation}
u = {1 / (2 {\rm sin}(\theta_j/2)) \over \jq + \jp \rm ln ( { {\rm csc} (\theta_0/2) + {\rm cot} (\theta_0/2)  \over {\rm csc} (\theta_j/2) + {\rm cot} (\theta_j/2) } )  }.
\end{equation}

However, the difference between this and the approximation (\ref{eqn:ut2}) is small, even for large opening angles, so (\ref{eqn:ut2}) will be used to describe the relationship between $u$ and $\theta_j$.

One final adjustment is made to (\ref{eqn:ut2}).  Numerical calculations have shown that as $\theta_j \rightarrow \pi/2$ (assuming a symmetric counter-jet), $u \rightarrow 0$.  In other words, the flow does not become perfectly spherical at a finite velocity (though it becomes nearly spherical at a finite but typically non-relativistic velocity).  To make the algebra consistent with this fact, a small constant term $u_0$ is subtracted from (\ref{eqn:ut2}):

\begin{equation}
u_0 \equiv {2 / \pi \over \jq + \jp \rm ln ( \pi / 2\theta_0) }
\end{equation}
so that
\begin{equation}
u = {1 / \theta_j \over \jq + \jp \rm ln (\theta_j/\theta_0) } - {2 / \pi \over \jq + \jp \rm ln ( \pi / 2\theta_0) }
\label{eqn:gamtheta}
\end{equation}

  To summarize, the jet evolution can be described by the following four equations:

\begin{equation}
u = {\Gamma_0 \over \sqrt{1 + 2 \Gamma_0 f + f^2} }\sqrt{1 + f/\Gamma_0}
\label{eqn:x}
\end{equation}
\begin{equation}
f = {\Omega \over M_0} \int_0^r \rho(r') r'^2 dr'
\end{equation}
\begin{equation}
\Omega = 4 \pi \rm sin^2(\theta_j/2)
\end{equation}
\begin{equation}
\theta_j = \left\{ \begin{array}{ll}
\theta_0     & \jq (u + u_0) \theta_0 < 1 \\
{ 1/(u+u_0) \over \jq + \jp {\rm ln}(\theta_j /\theta_0) } & {\rm otherwise}
\end{array} \right.
\label{eqn:y}
\end{equation}

Equation (\ref{eqn:y}) is solved iteratively for $\theta_j$, given $u$ (more concretely, equation (\ref{eqn:gamtheta}) is numerically solved for $\theta_j$ using Newton's method).  In practice, we will find the constants $\jq=1.6$ and $\jp=2.0$ result in an accurate solution for collision with a wind ($k=2$), while $\jq=2.5$, $\jp=4.0$ match well for a deceleration in a uniform density ($k=0$) medium.

\section{Numerical Set-Up} \label{sec:numerics}

\begin{figure}
\epsscale{1.15}
\plotone{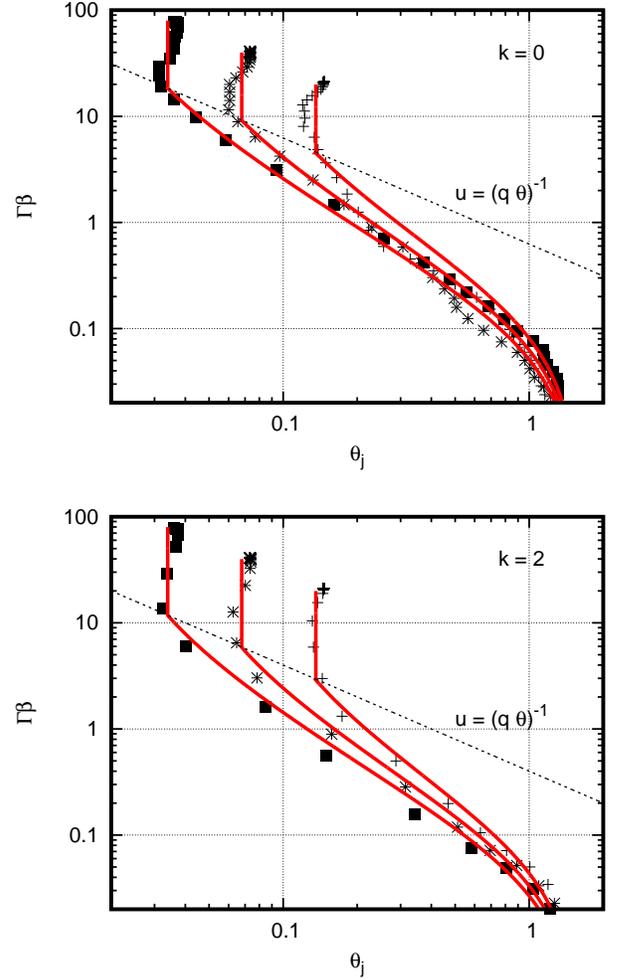}
\caption{ Investigation of the relationship between Lorentz factor and jet opening angle, for various initial conditions.  This relationship is well-fit by (\ref{eqn:y}).  Top Panel: jet spreading in the ISM (k=0).  Bottom Panel: jet spreading in a wind (k=2).
\label{fig:gamt} }
\end{figure}

\begin{figure}
\epsscale{1.15}
\plotone{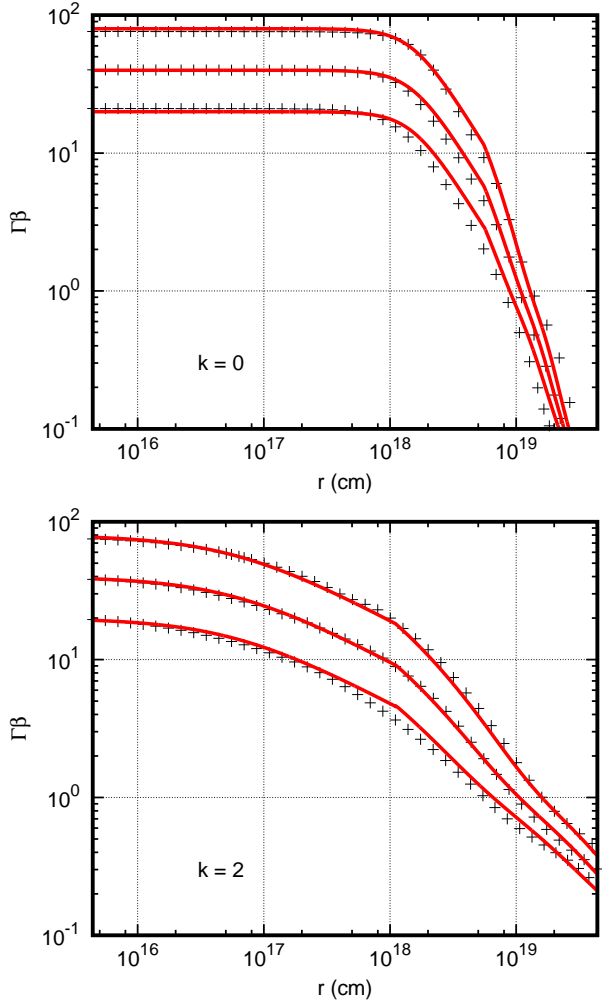}
\caption{ Evolution of Lorentz factor with time using (\ref{eqn:x}) to determine the spreading of the jet with opening angle.  Top Panel: jet spreading in the ISM (k=0).  Bottom Panel: jet spreading in a wind (k=2).
\label{fig:times} }
\end{figure}

\begin{figure}
\epsscale{1.15}
\plotone{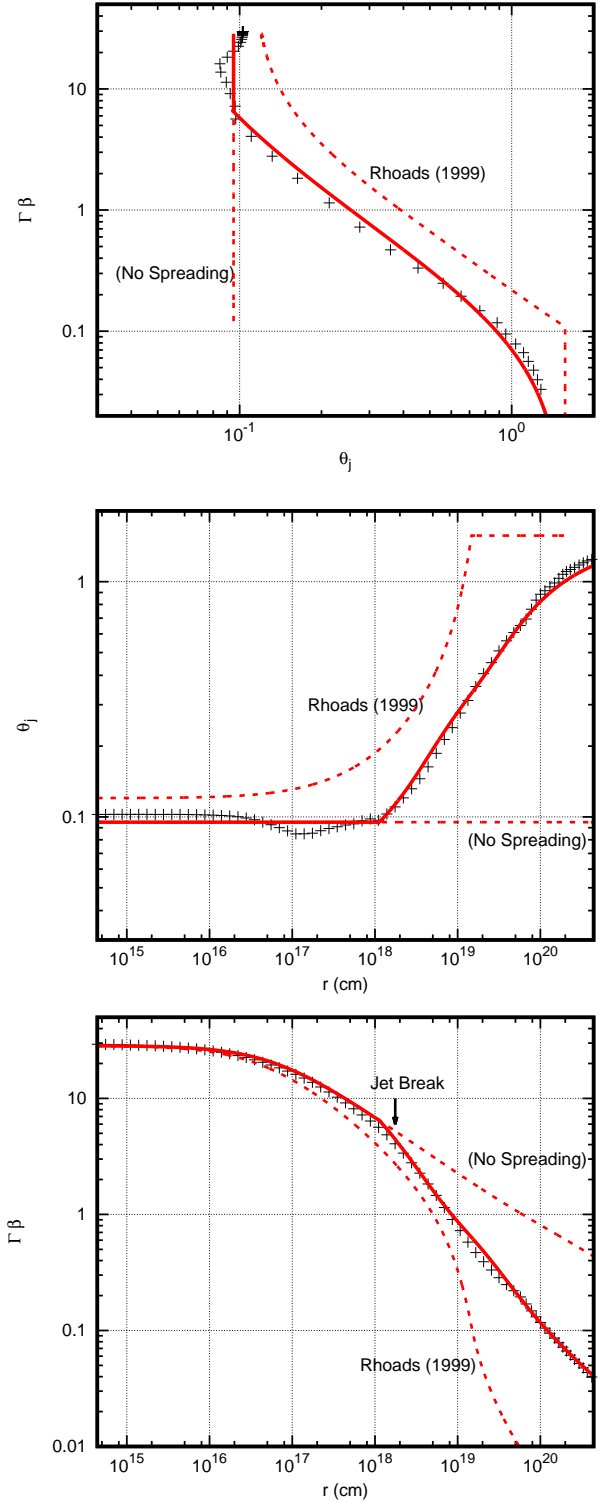}
\caption{ Improvement of the model over \cite{1999ApJ...525..737R}, for a jet spreading in a wind, with initial opening angle $\theta_0 = 0.1$.
\label{fig:rhoads} }
\end{figure}

The constants $\jp$ and $\jq$ above are measured from numerical calculations of jet spreading.  These numerical calculations are carried out using the JET code \citep{2011ApJS..197...15D, 2013ApJ...775...87D}.  the JET code solves the equations of relativistic gas dynamics using a moving mesh.  The mesh motion allows for evolution of very high Lorentz factors over many orders of magnitude of expansion.

\subsection{Initial Conditions}

Initial conditions are given by the ``boosted fireball" model of \cite{2013ApJ...776L...9D}.  The boosted fireball is parameterized by two Lorentz factors, $\eta_0$ and $\gamma_B$.  The model is found by taking a spherical explosion with Lorentz factor $\eta_0$ and boosting to a reference frame moving with relative Lorentz factor $\gamma_B$.  For ultra-relativistic Lorentz factors, this results in a flow with total Lorentz factor $\Gamma_0 \sim 2 \eta_0 \gamma_B$ and opening angle $\theta_0 \sim 1/\gamma_B$.
We fix $\eta_0 = 3$ and vary $\gamma_B$ in our setup, giving a flow with opening angle $1/\gamma_B$ and Lorentz factor $\approx 6 \gamma_B$, so that $\Gamma \theta_j \approx 6$ initially.

The jet propagates into the circumburst medium with density, $\rho = Ar^{-k}$.  In this study, we concentrate on the cases $k=2$ (wind) and $k=0$ (ISM).  The initial conditions are set at a very early time such that $c t = 10^{-6} (M_0/A)^{1 \over 3-k}$, where $M_0$ is the rest mass of the boosted fireball.

\subsection{Diagnostics}

To measure jet properties necessitates a reasonable definition for the opening angle $\theta_j$, to be measured from the numerical output.  Typically, numerical studies use $\theta_{90}$, or the opening angle containing $90\%$ of the jet's energy.  However, here we use an alternative definition, assuming that as the jet spreads, its energy is redistributed roughly evenly over the given solid angle.  From this, one can define the opening angle in terms of the isotropic equivalent energy:

\begin{equation}
\Omega = 4 \pi E / E_{\rm iso}.
\end{equation}

As $\Omega = 4 \pi ~ \rm sin^2( \theta_j / 2 )$,

\begin{equation}
{\rm sin}(\theta_j / 2) = \sqrt{ E / E_{\rm iso} }.
\end{equation}

$E_{\rm iso}$ is measured by an average of $4 \pi dE/d\Omega$ weighted by $dE/d\Omega$:

\begin{equation}
E_{\rm iso} = 4 \pi { \int (dE/d\Omega)^2 d\Omega \over \int (dE/d\Omega) d\Omega }.
\end{equation}

This results in the following formula for $\theta_j$, which is numerically evaluated at each time-step:

\begin{equation}
{\rm sin}(\theta_j/2) = { E \over \sqrt {4 \pi \int (dE/d\Omega)^2 d\Omega } }.
\end{equation}

The averaged four velocity is calculated by averaging over all cells, weighted by energy density:

\begin{equation}
\left<u\right> = {\int u(r,\theta) \tau dV \over \int \tau dV}.
\end{equation}

where $\tau = \gamma^2 \rho h - \gamma \rho$ is the energy density after subtracting rest mass energy density.  In this expression, $\gamma$ is the local fluid Lorentz factor, $\rho$ is proper density, and $h = 1 + e + P/\rho$ is the specific enthalpy, where $P$ is proper pressure and $e$ is specific internal energy.

\section{Results} \label{sec:dyn}

\subsection{Measurement of $u(\theta)$}

Four velocity and opening angle are measured at each timestep in the numerical calculation, and these values are plotted in Figure \ref{fig:gamt}.  Each jet traces out a curve through ($u$, $\theta$) space, each curve being determined by the initial opening angle, $\theta_0$.  These curves are fit to Equation (\ref{eqn:gamtheta}) in order to determine the constants $\jp$ and $\jq$.  The dynamics are reasonably well-fit by $\jp = 2.0$, $\jq = 1.6$, when the jet is colliding with a wind ($\rho \propto r^{-2}$).  When the surrounding medium is of uniform density ($\rho = $const.), the solution is consistent with $\jp = 4.0$, $\jq = 2.5$.  

\subsection{Jet Deceleration}

Using the analytical equations for $u$ and $\theta_j$ (\ref{eqn:x}-\ref{eqn:y}) along with the appropriate choice of $\jp$ and $\jq$, it is now possible to accurately model the evolution of the Lorentz factor with time.  Figure \ref{fig:times} shows this evolution for the various initial jet models, showing that the model can accurately reproduce deceleration and spreading of the jet.  The deceleration process is recovered in both the $k=0$ (ISM) and $k=2$ (wind) cases.

Figure \ref{fig:rhoads} shows that this is a substantial improvement to the model of \cite{1999ApJ...525..737R}.  Using an example with initial opening angle of $\theta_0 \approx 0.1$ (corresponding to $\eta_0 = 3$, $\gamma_B = 10$) colliding with a wind ($k=2$), the opening angle and four velocity as a function of time are compared with the fitted model, alongside the Rhoads model, and a model which ignores the jet's spreading, keeping the opening angle fixed at $\theta_j = \theta_0$.  The model is able to accurately reproduce the evolution of $\theta_j$ and $u$ with time, much better than the Rhoads model.

\section{Discussion} \label{sec:disc}

We have presented a new model for the dynamics of a decelerating and spreading relativistic jet.  The model is built from dynamical considerations, and matched to numerical results via two fitting parameters.  By performing a survey over initial opening angle $\theta_0$ and density power-law index $k$ we have shown that the model is broadly effective over the relevant parameter space of GRB and TDE jets.

The fitted model has been demonstrated to be an improvement to the original model of \cite{1999ApJ...525..737R}, and is also an improvement to the model of \cite{2012MNRAS.421..570G}, as the model is tailored to match with numerical calculations, and is consistent over a significant range of parameter space, including narrow initial opening angles $\theta_0 \sim 0.1$ relevant for gamma ray bursts.

The model can be used to construct synthetic light curves and spectra for TDEs and GRB afterglows.  This will be the focus of paper II.  Because the jet model is analytical, these synthetic observations can be generated rapidly, making it possible to find best-fit jet parameters to observations.

\acknowledgments

TL is a Jansky Fellow of the National Radio Astronomy Observatory. This work was supported in part by the Theoretical Astrophysics Center at UC Berkeley and by the Gordon and Betty Moore Foundation through Grant GBMF5076.  Early stages of research used the Savio computational cluster resource provided by the Berkeley Research Computing program at the University of California, Berkeley (supported by the UC Berkeley Chancellor, Vice Chancellor of Research, and Office of the CIO).  High-resolution calculations were provided by the NASA High-End Computing (HEC) Program through the NASA Advanced Supercomputing (NAS) Division at Ames Research Center.  

\bibliographystyle{apj} 

\begin{thebibliography}{}
\expandafter\ifx\csname natexlab\endcsname\relax\def\natexlab#1{#1}\fi

\bibitem[{Cenko {et~al.}(2010)Cenko, Frail, Harrison, Kulkarni, Nakar, Chandra,
  Butler, Fox, Gal-Yam, Kasliwal, Kelemen, Moon, Ofek, Price, Rau, Soderberg,
  Teplitz, Werner, Bock, Bloom, Starr, Filippenko, Chevalier, Gehrels, Nousek,
  \& Piran}]{cfh+10}
Cenko, S.~B., Frail, D.~A., Harrison, F.~A., {et~al.} 2010, \apj, 711, 641

\bibitem[{Cenko {et~al.}(2011)Cenko, Frail, Harrison, Haislip, Reichart,
  Butler, Cobb, Cucchiara, Berger, Bloom, Chandra, Fox, Perley, Prochaska,
  Filippenko, Glazebrook, Ivarsen, Kasliwal, Kulkarni, LaCluyze, Lopez, Morgan,
  Pettini, \& Rana}]{cfh+11}
---. 2011, \apj, 732, 29

\bibitem[{{De Colle} {et~al.}(2012){De Colle}, {Ramirez-Ruiz}, {Granot}, \&
  {Lopez-Camara}}]{2012ApJ...751...57D}
{De Colle}, F., {Ramirez-Ruiz}, E., {Granot}, J., \& {Lopez-Camara}, D. 2012,
  \apj, 751, 57

\bibitem[{{Duffell} \& {MacFadyen}(2011)}]{2011ApJS..197...15D}
{Duffell}, P.~C., \& {MacFadyen}, A.~I. 2011, \apjs, 197, 15

\bibitem[{{Duffell} \& {MacFadyen}(2013{\natexlab{a}})}]{2013ApJ...776L...9D}
---. 2013{\natexlab{a}}, \apjl, 776, L9

\bibitem[{{Duffell} \& {MacFadyen}(2013{\natexlab{b}})}]{2013ApJ...775...87D}
---. 2013{\natexlab{b}}, \apj, 775, 87

\bibitem[{{Duffell} \& {MacFadyen}(2015)}]{2015ApJ...806..205D}
---. 2015, \apj, 806, 205

\bibitem[{{Granot} \& {Piran}(2012)}]{2012MNRAS.421..570G}
{Granot}, J., \& {Piran}, T. 2012, \mnras, 421, 570

\bibitem[{{Gruzinov}(2007)}]{2007arXiv0704.3081G}
{Gruzinov}, A. 2007, ArXiv e-prints, arXiv:0704.3081

\bibitem[{{Keshet} \& {Kogan}(2015)}]{2015ApJ...815..100K}
{Keshet}, U., \& {Kogan}, D. 2015, \apj, 815, 100

\bibitem[{{Laskar} {et~al.}(2015){Laskar}, {Berger}, {Margutti}, {Perley},
  {Zauderer}, {Sari}, \& {Fong}}]{2015ApJ...814....1L}
{Laskar}, T., {Berger}, E., {Margutti}, R., {et~al.} 2015, \apj, 814, 1

\bibitem[{{Lyutikov}(2012)}]{2012MNRAS.421..522L}
{Lyutikov}, M. 2012, \mnras, 421, 522

\bibitem[{{Nousek} {et~al.}(2006){Nousek}, {Kouveliotou}, {Grupe}, {Page},
  {Granot}, {Ramirez-Ruiz}, {Patel}, {Burrows}, {Mangano}, {Barthelmy},
  {Beardmore}, {Campana}, {Capalbi}, {Chincarini}, {Cusumano}, {Falcone},
  {Gehrels}, {Giommi}, {Goad}, {Godet}, {Hurkett}, {Kennea}, {Moretti},
  {O'Brien}, {Osborne}, {Romano}, {Tagliaferri}, \&
  {Wells}}]{2006ApJ...642..389N}
{Nousek}, J.~A., {Kouveliotou}, C., {Grupe}, D., {et~al.} 2006, \apj, 642, 389

\bibitem[{{Paczynski} \& {Rhoads}(1993)}]{1993ApJ...418L...5P}
{Paczynski}, B., \& {Rhoads}, J.~E. 1993, \apjl, 418, L5

\bibitem[{{Racusin} {et~al.}(2009){Racusin}, {Liang}, {Burrows}, {Falcone},
  {Sakamoto}, {Zhang}, {Zhang}, {Evans}, \& {Osborne}}]{2009ApJ...698...43R}
{Racusin}, J.~L., {Liang}, E.~W., {Burrows}, D.~N., {et~al.} 2009, \apj, 698,
  43

\bibitem[{{Rhoads}(1999)}]{1999ApJ...525..737R}
{Rhoads}, J.~E. 1999, \apj, 525, 737

\bibitem[{{Ryan} {et~al.}(2015){Ryan}, {van Eerten}, {MacFadyen}, \&
  {Zhang}}]{2015ApJ...799....3R}
{Ryan}, G., {van Eerten}, H., {MacFadyen}, A., \& {Zhang}, B.-B. 2015, \apj,
  799, 3

\bibitem[{{Sari} {et~al.}(1999){Sari}, {Piran}, \&
  {Halpern}}]{1999ApJ...519L..17S}
{Sari}, R., {Piran}, T., \& {Halpern}, J.~P. 1999, \apjl, 519, L17

\bibitem[{{van Eerten} \& {MacFadyen}(2013)}]{2013ApJ...767..141V}
{van Eerten}, H., \& {MacFadyen}, A. 2013, \apj, 767, 141

\bibitem[{{van Eerten} {et~al.}(2012){van Eerten}, {van der Horst}, \&
  {MacFadyen}}]{2012ApJ...749...44V}
{van Eerten}, H., {van der Horst}, A., \& {MacFadyen}, A. 2012, \apj, 749, 44

\bibitem[{{van Eerten} {et~al.}(2010{\natexlab{a}}){van Eerten}, {Zhang}, \&
  {MacFadyen}}]{2010ApJ...722..235V}
{van Eerten}, H., {Zhang}, W., \& {MacFadyen}, A. 2010{\natexlab{a}}, \apj,
  722, 235

\bibitem[{{van Eerten} {et~al.}(2010{\natexlab{b}}){van Eerten}, {Leventis},
  {Meliani}, {Wijers}, \& {Keppens}}]{2010MNRAS.403..300V}
{van Eerten}, H.~J., {Leventis}, K., {Meliani}, Z., {Wijers}, R.~A.~M.~J., \&
  {Keppens}, R. 2010{\natexlab{b}}, \mnras, 403, 300

\bibitem[{{van Eerten} \& {MacFadyen}(2011)}]{2011ApJ...733L..37V}
{van Eerten}, H.~J., \& {MacFadyen}, A.~I. 2011, \apjl, 733, L37

\bibitem[{{van Eerten} \&
  {MacFadyen}(2012{\natexlab{a}})}]{2012ApJ...747L..30V}
---. 2012{\natexlab{a}}, \apjl, 747, L30

\bibitem[{{van Eerten} \&
  {MacFadyen}(2012{\natexlab{b}})}]{2012ApJ...751..155V}
---. 2012{\natexlab{b}}, \apj, 751, 155

\bibitem[{{Zhang} {et~al.}(2015){Zhang}, {van Eerten}, {Burrows}, {Ryan},
  {Evans}, {Racusin}, {Troja}, \& {MacFadyen}}]{2015ApJ...806...15Z}
{Zhang}, B.-B., {van Eerten}, H., {Burrows}, D.~N., {et~al.} 2015, \apj, 806,
  15

\bibitem[{{Zhang} \& {MacFadyen}(2009)}]{2009ApJ...698.1261Z}
{Zhang}, W., \& {MacFadyen}, A. 2009, \apj, 698, 1261

\end{thebibliography}

\end{document}